# A QUANTITATIVE APPROACH IN HEURISTIC EVALUATION OF E-COMMERCE WEBSITES

Xiaosong Li, Ye Liu, Zizhou Fan and Will Li

Computer Science Practice Pathway,
Unitec Institute of Technology, Auckland, New Zealand

## ABSTRACT

*This paper presents a pilot study on developing an instrument to predict the quality of e-commerce websites. The 8C model was adopted as the reference model of the heuristic evaluation. Each dimension of the 8C was mapped into a set of quantitative website elements, selected websites were scraped to get the quantitative website elements, and the score of each dimension was calculated. A software was developed in PHP for the experiments. In the training process, 10 experiments were conducted and quantitative analyses were regressively conducted between the experiments. The conversion rate was used to verify the heuristic evaluation of an e-commerce website after each experiment. The results showed that the mapping revisions between the experiments improved the performance of the evaluation instrument, therefore the experiment process and the quantitative mapping revision guideline proposed was on the right track. The software resulted from the experiment 10 can serve as the aimed e-commerce website evaluation instrument. The experiment results and the future work have been discussed.*

## 1. INTRODUCTION

E-commerce websites have been increased greatly in the new era; they face many competitors. Research revealed that efforts put into usability design and modification improved the performance of usability on websites greatly [1]. To help website developers and other stakeholders understand how to develop e-commerce websites properly and maximize profit, many evaluation methods have been developed [1, 2]. One approach is called user based testing [1], which takes into account subjective perception, both in terms of website content and design. This perception varies with the expertise, the cognitive skills and the end goal of each user [1]. If an automatic approach is used to evaluate website content and design from the user's perspective, that should standardize the evaluation process and make the evaluation consistent and objective.

7C framework was introduced to evaluate the quality of e-commerce website content and user interface design [3], which is considered as a useful reference model for developers, analysts, managers, and executives, when designing and/or evaluating the interface channels between the customer and the web based application. However, it is insufficient to completely address the new generation of web applications [4]. Collaboration and user-generated content are important features in the new generation websites. The 7C framework was extended into the 8C framework by adding collaboration as the 8th element in the model and the meaning of each of the eight design elements was updated as well, so that they are effective in representing the interface design elements of new generation websites [4].

This paper presents a pilot study on developing a software instrument to predict the quality of e-commerce websites. The objective of the resulting instrument is to provide a meaningful estimation on the quality of a given e-commerce website. The 8C model was adopted as the reference model of the heuristic evaluation. Each dimension of the 8C model was mapped into a formula consisting of a set of quantitative website elements, the websites were scraped to get the quantitative website elements, and the score of each dimension was calculated based on the formula. Another formula was defined to calculate the total score for the website based on the scores from each dimension.

A software was developed in PHP for both training and testing experiments. An experimental process and its quantitative mapping revision guideline were proposed and used. In the training process, 10 experiments were conducted and quantitative analyses were conducted regressively between the experiments. The conversion rate was used in this study to test and verify the heuristic evaluation of an e-commerce website. 100 websites from five different categories were selected as the training data. 7 websites ordered by the conversion rate were used as testing data to test the results at the end of each experiment in the training process and 15 websites ordered by the CR were used as the verifying data at the end of all the experiments.

This paper presents the following order: Section 2 describes the related work and the method used; Section 3 explains the design of the experiments completed in this study. Section 4 describes the details of the experiments conducted and the results are presented and discussed. Lastly, Section 5 summarises the work presented and introduces future work.

## 2. THE RELATED WORK AND THE METHOD

The heuristic evaluation method is a technique for evaluating the usability, with the inspection being carried out mainly by evaluators from principles established by the discipline [5]. In most applications the results tend to be qualitative, however, these qualitative results do not allow us to determine how usable it is or how it becomes an interactive system. Hence, the need for quantitative results may also be very necessary in order to determine the effort that would be needed to get a sufficiently usable system [5].

Usually when a website is evaluated against the 7C framework, similar heuristic evaluation method is used, where subjective perception is applied. For example, in [6], to evaluate a group of 4 and 5-star luxury hotel websites against the 7C framework, a checklist consisting of 63 checkpoints was developed based on research literature and expert opinions. This approach again could be inconsistent and subjective. An automatic approach could standardize this evaluation process and make it more consistent and objective.

The accurate prediction of a numerical target variable is an important task in machine learning. Quantitative heuristic analysis has been used in machine learning to predict various values in the data mining and inductive rule learning communities, where a strong focus lies on the comprehensibility of the learned models [7]. In [7], a heuristic rule learning algorithm that learns regression models is used where, a region around the target value predicted by the rule is dynamically defined. In [8], a unified measure of web usability was used based on a multiple regression model, and then the estimated index is used to measure its impact on community bank performance. Results showed that banks with higher usability score perform significantly better than those with lower score.

To automise the process of the heuristic evaluation, quantitative measurements should be used to determine the effort that would be needed to get a sufficiently usable system [5]. The quantitative measurements can be used to indicate how effective an e-commerce website is. In a project described in [5], a function named USABAIPO-H was defined, the purpose was to process quantitative results of the heuristic evaluation. They suggested that a framework should integrate effective tools for measuring the usability level of the developed system. The framework referred to a necessary experiment carried out that will enable the framework development. To accomplish that, and with the aim of incorporating metrics related with heuristic evaluation, they have prompted a new experimental work to check the validity of the above mentioned experiment and, if appropriate, correct or improve the previously obtained results. They have done a quantitative analysis of the obtained results. Once each website is analysed and after applying a modified formula they introduced, the usability percentage is obtained corresponding to each website to give estimation on the usability level of the selected websites. Similar approach and process (framework) were used in our study.

Heuristic evaluation is done as a systematic inspection of a user interface design for usability. The goal of this is to find the usability problems in the design so that they can be attended to as part of an iterative design process [5, 9]. This was done slightly differently in our study, quantitative measurements defined and implemented in our study were trying to measure how well a website complied with the evaluation principles used. Of course, those with low scores indicate shortages in some directions.

In [5], the steps used by Nielsen in the heuristic evaluation [10] had been summarised as the following, which was integrated with the severity ratings found in [11]. It was recommended that these steps should be followed to make a heuristic evaluation efficient and to provide quality results [5, 10 and 11].

1. **Prior training**: The evaluator must become familiar with the interface for a few minutes to learn the website and to be able to carry out the heuristic evaluation agilely.

2. **Evaluations**: The evaluator follows the set of heuristics to find deficiencies or to catalogue the website as usable. He can write comments.

3. **Rate severity**: The evaluator should determine the severity of each of the problems encountered. It is therefore appropriate that the importance of the problems is rated. The three usability rating factors introduced in [11] were adopted as parameters to indicate how severe a problem found during the evaluation, including the **frequency** with which problems occur (common or rare?); the **impact** of the problem, if users are very affected when this happens; and the **persistence** of the problem. Is it a one-time problem that users can overcome once they know about it or will users repeatedly be bothered by the problem?

4. **Review**: To analyse each of the evaluations made to present a report with all the problems and possible resolutions; taking into account the qualitative analysis obtained.

The above steps made it easy for automatic heuristic evaluation, which were adopted in our e-commerce website evaluation, where human experts were replaced by a self-developed software instrument.

Prior training is very import. Instead of human experts, we need to train our software instrument to become an expert in the e-commerce website evaluation. In this step, simple machine learning techniques were used. 100 websites from five different categories (*Electronics*, *Publishing & entertainment*, *Home and garden*, *Books*, *Industrial equipment*), 20 from each category were selected as the training data. The initial software instrument was used to evaluate all the training websites in the first experiment, the results were tested against the conversion rate (CR). The software was improved based on the test results. This process was repeated for 10 experiments, that is, a regression process was carried to refine (train) the software instrument, where 10 experiments were used for the training. Conversion rate (CR) was used to test the results after each experiment. The resulting software instrument from 10[th] experiment can be used for the future e-commerce websites evaluation.

For the evaluation step, this study considered the seven dimensions defined in the 7C framework and the additional dimension "collaboration" introduced in the 8C framework. For the web 2.0 features, only those features easy to be obtained via web scraping were considered such as website forum, blog and Ajax. Table 1 presents the key meaning of each dimension in 8C [4]. The 8 dimensions introduced in the 8C framework were used as a set of heuristics (evaluation principles) in our study. Each dimension was mapped into a set of quantitative website elements. The details are given in Section 3.

Table 1. The key meaning of each dimension in 8C.

| Dimensions | Meanings |
|---|---|
| 1: Context | How the site is organized, and how the content is presented to the users? |
| 2: Content | What are offered by the site? |
| 3: Community | Non-interactive communication; Interactive communication. |
| 4: Customization | Refers to the site's ability to tailor itself (tailoring) or to be tailored. |
| 5: Communication | Site-to-user communications. |
| 6: Connection | Refers to the extent of formal linkage from one site to others. |
| 7: Commerce | Deals with the interface that supports the various aspects of e-commerce. |
| 8: Collaboration | Generally in the form of feedback forms, forums, and bulletin boards. |

Conversion rate (CR) is the percentage of users who take a desired action. The typical example of conversion rate is the percentage of website visitors who buy something on the site, For the purpose of managing user interface design and tracking the effectiveness of user experience efforts, the conversion rate is usually very important [12]. The conversion rate measures what happens once people are at your website, which is greatly influenced by the design and is a key parameter to track for assessing whether a user experience strategy is working. Lower conversion rates? You must be doing something wrong with the design. Higher conversion rates? You can praise your designers [12]. This suggests that there is a proportional relationship between the conversion rate of an e-commerce website and its user interface design. It is reasonable to use the conversion rate to measure the quality of the user interface of an e-commerce website.

Measuring the user experience offers so much more than just simple observation. Metrics add structure to the design and evaluation process, give insight into the findings and provide information to the decision makers [13]. The five categories were selected from [14], where the CRs for 25 retail categories were listed. **Electronics** and **Publishing & entertainment** were associated with high level CR; **Home & garden** and **Books** were associated with middle level CR; and **Industrial equipment** were associated with low level CR.

In the rating step, the score of each dimension was calculated first and then the score for the whole website was generated based on that. The calculations were based on the quantitative measurements of the website properties. The frequencies of the properties were included as well. The details and the formula are given in Section 3. The score of each dimension indicates how well the evaluated website meets with the requirements of that dimension. Low score indicates shortages in the dimension.

For the review step, after the evaluation of a website, a summary of the scores achieved for each 8C dimension were saved into an Excel file and the total score for the evaluated website was saved as well. This information will be used by the users of the software instrument for comparison and analysis purposes.

The process for our automatic heuristic evaluation of the e-commerce websites by using the final software instrument developed for this research is as the following:

1. Start the final software instrument.

2. Input the URL for the website to be evaluated.

3. Run the software instrument.

4. Download the output Excel file.

5. Analyse the scores from the output file.

## 3. THE EXPERIMENT ENVIRONMENT AND DESIGN

Quantitative usability estimation is typically associated with the calculation of metrics that assess dimensions of software quality [5].

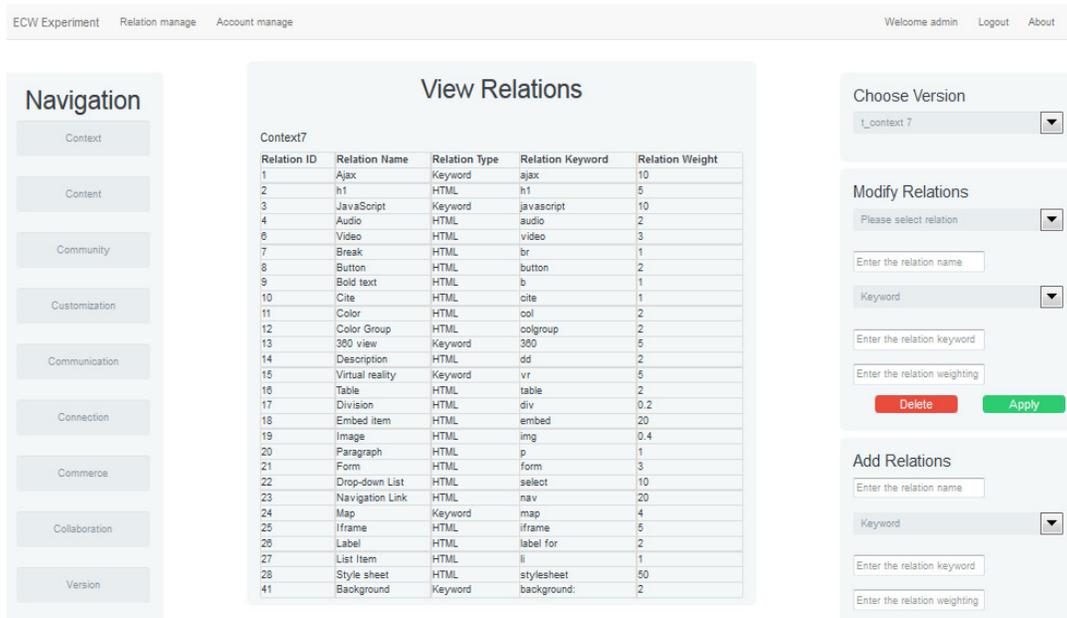

Figure 1. The mapping management UI and the relations in a mapping

A software written in PHP is a web application, which was developed for both training and testing experiments. Figure 1 shows the software mapping management user interface with the mapping relations between the **Context** dimension and the selected HTML tags/keywords in **Experiment 7**. For an e-commerce website to be experimented (evaluated), only the home page was considered in this study.

The software supports two user roles: an administrator user and a researcher user. The administrator can create or modify the mappings between the website quantitative elements and the dimensions of 8C. The researchers can only do experiments (evaluating the given websites). The administrator can do the experiments as well. The software records each group of mappings as an experiments. A user can select any predefined experiment to perform website evaluation.

This was designed and implemented for the evaluation instrument training. The software can evaluate either a single website or a set of websites at once. For a single website, the URL of the website can be either manually typed in from an input text box or uploaded from a text file. For a set of websites, the URLs can be uploaded from a text file. The resulting scores are saved into an Excel file which can be downloaded from the web application.

Two major approaches were used to identify the website quantitative elements and calculate the metrics for each dimension: finding keywords and scraping HTML tags, where a keyword could be an important text or a JavaScript/CSS keyword. Each keyword or HTML tag is associated with a numeric weight, which determines the importance of the relation, higher weight means more important. The mapping relations between each dimension and the selected keyword or HTML tag are defined before each experiment, which can be adjusted in the subsequent experiments based on the experiment results.

Let $NRj$ be the total number of the relations in a mapping between dimension j and the selected HTML tags/keywords; $RSi$ be the score of relation $i$; $Wi$ be the associated weight of relation $i$; if the relation $i$ is a keyword, $RSi$ will be $Wi$; if relation $i$ is an HTML tag, $RSi$ will be calculated by the following formula:

$$RSi = \frac{STagNi}{TTagNi} * Scalar * Wi \qquad (1)$$

Where $STagNi$ is the number of the occurrence of the selected HTML tag for relation $i$; $TTagNi$ is the total number of HTML tag on the selected page; Scalar is set as 100 to make the score a meaningful magnitude. The total score $TS$ is the sum of the scores for all 8 dimensions in 8C framework.

$$TS = \sum_{j=1}^{8} \sum_{i=1}^{NRj} (RSi) \qquad (2)$$

An experimental process and its revision guideline were proposed and used. Initially, in **Experiment 1**, only the keywords/HTML tags that can intuitively reflect the meaning of a dimension as defined in the 8C framework were selected as the relations for the mapping of that dimension heuristically. The weights for the relations also were selected in the similar way heuristically.

Then the scores for all the training websites were calculated respectively according to formula (2). The training websites were ordered based on their CRs (CR) first, and then the training websites were ordered again based on their scores. If the score order is different from the CR order, the mappings for all the 8 dimensions were reviewed and revised in the following three aspects:

1. Check if any relation score is dominating the dimension score based on the overall performance of the training websites, if yes, adjust the weight of that relation to make the relation score of a meaningful magnitude.

2. Check if the score of any dimension is dominating the total score based on the overall performance of the training websites, if yes, scale all the scores in that dimension to make the dimension score of a meaningful magnitude.

3. Recheck all the mappings against the 8C model and make adjustment accordingly. This may involve adding or deleting relations.

The above would result in the new mappings for the next experiment. This process went through regressively for 10 experiments. As an example, Table 2 shows the mappings for *Collaboration* dimension in *Experiment 1*, *Experiment 6* and *Experiment 8*.

Table 2.  The mapping for Collaboration in three experiments.

| Experiment 1 | | Experiment 6 | | Experiment 8 | |
|---|---|---|---|---|---|
| Relation Name | Relation Weight | Relation Name | Relation Weight | Relation Name | Relation Weight |
| Forums | 3 | Forums | 3 | Forums | 3 |
| Bulletin boards | 3 | Bulletin boards | 3 | Bulletin boards | 3 |
| FAQ | 3 | FAQ | 3 | FAQ | 3 |
| | | Feedback | 5 | Feedback | 5 |
| | | | | Review | 5 |
| | | | | Suggestion | 5 |
| | | | | Comment | 5 |

## 4. THE EXPERIMENT RESULTS AND DISCUSSION

100 websites from five different categories (Electronics, Publishing & entertainment, Home and garden, Books, Industrial equipment), 20 from each category were selected as the training data. The five categories were selected from [14], where the CRs for 25 retail categories were listed. *Electronics* and *Publishing & entertainment* were associated with high level CR; *Home & garden* and *Books* were associated with middle level CR; and *Industrial equipment* were associated with low level CR.

Table 3.  The categories of training data.

| Categories | Conversion Rates |
|---|---|
| Electronics | Around 23% |
| Publishing & entertainment | Around 20% |
| Home & garden | Around 14% |
| Books | Around 13% |
| Industrial equipment | Around 7% |

The top 10 e-commerce websites based on CR for 2010 were listed in [15], only 7 of them were valid for the experiments, and they all were used to test the results at the end of each experiment for all the 10 experiments. Table 4 shows the 7 testing websites.

Table 4.  The testing data.

| Website Names | Conversion Rates |
|---|---|
| Woman Within | 25.3% |
| Blair | 20.4% |
| 1800petmeds | 17.7% |
| qvc | 16% |
| ProFlowers | 15.8% |
| Oriental Trading Company | 14.9% |
| Roamans | 14.4% |

After each experiment, the training websites were ordered again based on their scores. If the score order is different from the CR order, the mappings for all the 8 dimensions in 8C model were reviewed and then revised if needed, this resulted in the new mapping for the next experiment. Figure 2 shows the absolute score differences between the expected order and the actual order. [12] suggests that there is a proportional relationship between the CR of an e-commerce website and its user interface design. It is reasonable to assume that the less the difference, the more accurate the evaluation.

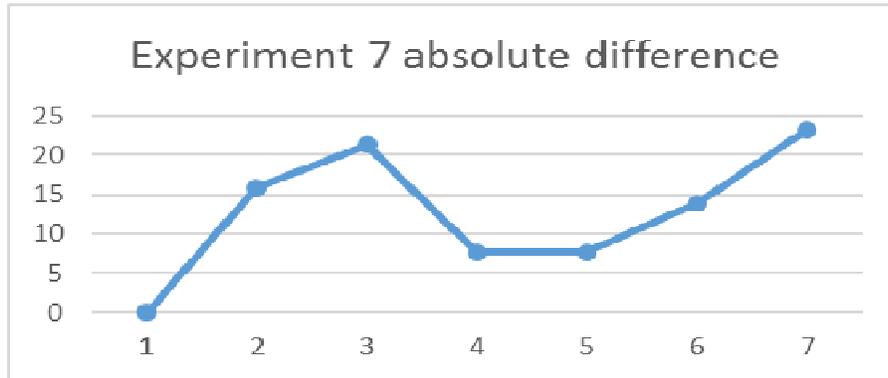

Figure 2. The absolute score differences between the expected order and the actual order for one experiment

The differences in each experiment for all the 7 training websites were averaged and Figure 3 shows the average for all the experiments except **Experiment 9**. As the scores of **Experiment 10** were obtained by scaling the scores in **Experiment 9** by 10%. It was observed that the trends of the curve going down along the experiments. This suggested that the mapping revisions between the experiments improved the performance of the evaluation instrument and it is positive.

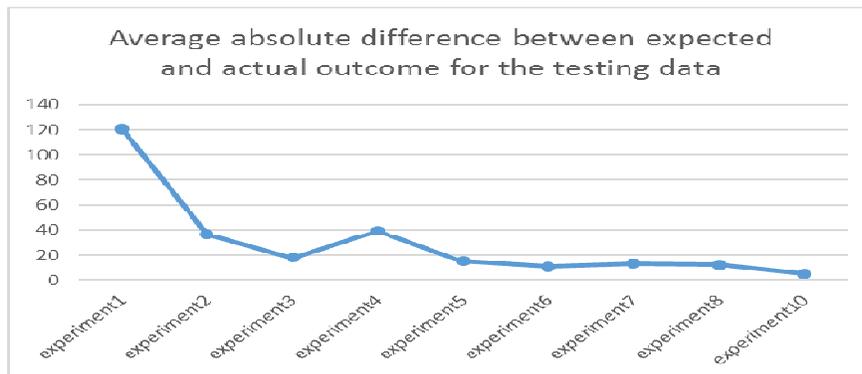

Figure 3. The average absolute difference between expected and actual outcomes for nine experiments

Table 5. The dimension contribution analysis of Experiment 8 & 10.

| Attribute Category | Experiment 8 | | | Experiment 10 | | |
|---|---|---|---|---|---|---|
| | Attribute Number | Contribution to the total score | Standard deviation | Attribute Number | Contribution to the total score | Standard deviation |
| Context | 28 | 40.94% | 15.43 | 28 | 14.71% | 1.55 |
| Content | 11 | 5.93% | 7.08 | 11 | 8.21% | 2.62 |
| Community | 18 | 9.04% | 7.77 | 18 | 12.60% | 3.13 |
| Customization | 5 | 7.85% | 6.00 | 5 | 11.22% | 2.43 |
| Communication | 13 | 12.47% | 6.55 | 13 | 17.79% | 2.65 |
| Connection | 2 | 5.15% | 4.81 | 2 | 7.44% | 1.97 |
| Commerce | 10 | 14.68% | 8.30 | 10 | 15.80% | 2.55 |
| Collaboration | 7 | 3.93% | 3.94 | 7 | 12.22% | 3.48 |

In **Experiment 8**, it was observed that some of the dimensions' scores dominated the total score of the website. Table 5 shows the dimension contribution analysis of **Experiment 8 & 10**, where the number of attributes number is the number of relations in the mapping for each dimension (Attribute Category) of the 8C; contribution to the total score is the sum of the scores in a dimension for all the training websites divided by the total score of all the training websites in an experiment. **Context** made much more contribution (40.94%) than the others did. On the other hand, some were too small to influence the total score, such as Content (5.93%) **Connection** (5.93%) and **Collaboration** (5.93%).

The standard deviation can provide some ideas on whether the attributes in a dimension is informative. For example, standard deviation for **Collaboration** was the smallest one in **Experiment 8**, however, there were 7 attributes in this dimension. This suggested that the meaning of the attributes might be overlapping. So standard deviation for each dimension over all the training websites should be considered in the review process after each experiment in the future study.

In this study, scaling the scores for the dimensions were attempted to balance the influences of all the dimensions. For a website, let TS be its total score, and let score codes and scale parameter codes be defined in Table 6.

Table 6. Codes used in the scale formula.

| Score Code | Meaning of the code | Scale Parameter | Scale Number |
|---|---|---|---|
| SC1 | Score of Context | P1 | 1 |
| SC2 | Score of Content | P2 | 4 |
| SC3 | Score of Community | P3 | 4 |
| SC4 | Score of Customization | P4 | 4 |
| SC5 | Score of Communication | P5 | 4 |
| SC6 | Score of Connection | P6 | 4 |
| SC7 | Score of Commerce | P7 | 3 |
| SC8 | Score of Collaboration | P8 | 9 |

Formula (3) was used to calculate *TS* in **Experiment 9**, the resulting scores were much larger than the other experiments, so the results were divided by 10 for further scaling, which were recorded as **Experiment 10**.

$$TS = \sum_{i=1}^{8} (SCi * Pi) \qquad (3)$$

The right side of Table 5 shows the contribution of each dimension after the scaling in **Experiment 10**. This time the contributions of the dimensions are much balanced.

The verifying data was obtained from [16], which listed top 15 e-commerce websites based on CR for 2014. All of them were valid for the experiments and were used to check the mappings used in all the experiments except **Experiment 9** as **Experiment 10** can represent **Experiment 9**.

Table 7 shows the order of the 15 verifying websites. Figure 4 shows the average absolute difference between expected and actual outcomes for the 15 verifying websites. It was observed that the trend of the curve was going down along the experiments, which was consistent with the testing results of Figure 3. This suggested that the experiments were on the right track and the results were positive. The resulting instrument from **Experiment 10** could be used to evaluate a given e-commence website and provide meaningful estimation on the quality of the website.

Table 7.  The 2014 data.

| Website Names | Conversion Rates |
|---|---|
| Play.Google | 30.00% |
| MovieMars | 22.95% |
| DollarShaveClub | 20.00% |
| 1800Contacts | 18.40% |
| 1800Flowers | 16.90% |
| Coastal | 14.50% |
| Keurig | 13.00% |
| FTD | 11.70% |
| ProFlowers | 11.70% |
| PureFormulas | 10.74% |
| FreshDirect | 10.50% |
| TheGreatCourses | 10.04% |
| 1800PetMeds | 10.00% |
| AmeriMark | 10.00% |
| OvernightPrints | 9.95% |

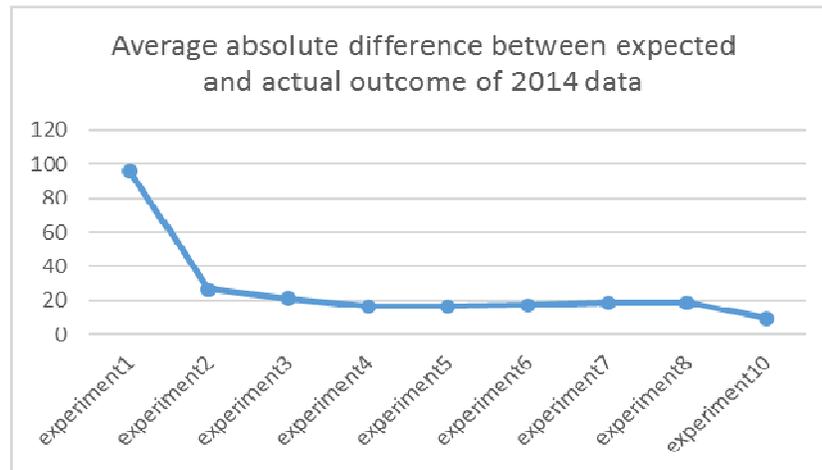

Figure 4.  The average absolute difference between expected and actual outcomes for 2014 data.

Table 8.  The average score of each category in each experiment.

| No | Electronics | Entertainment | Home | Books | Industrial |
|---|---|---|---|---|---|
| 1 | 165.27 | 142.92 | 157.36 | 152.41 | 148.19 |
| 2 | 144.16 | 124.11 | 124.71 | 129.84 | 124.40 |
| 3 | 108.77 | 83.19 | 85.73 | 93.20 | 88.15 |
| 4 | 98.73 | 85.15 | 90.07 | 91.09 | 94.00 |
| 5 | 90.08 | 77.28 | 82.28 | 87.61 | 83.50 |
| 6 | 91.27 | 81.50 | 84.73 | 86.80 | 89.00 |
| 7 | 163.41 | 141.10 | 145.65 | 136.69 | 148.75 |
| 8 | 159.83 | 138.54 | 142.54 | 134.37 | 145.52 |
| 9 | 467.53 | 395.98 | 380.85 | 358.28 | 403.04 |
| 10 | 46.77 | 39.89 | 40.21 | 36.00 | 40.34 |

Table 8 shows the average score of each category in each experiment. According to Table 3, websites in **Electronics** category should have the highest scores; websites in **Industrial Equipment** category should have the lowest scores; and **Books** are in the middle. In Table 8, the *Electronics* websites always have the highest score in all the experiments, *Books* websites are in the middle sometimes, particularly in *Experiment 10*. These are consistent between the two tables (Table 3 and Table 8). However, *Industrial Equipment* websites usually do not have the lowest scores. This suggests that the website design and usability could have an impact on an e-commerce website's CR, however, there are other factors as well, such as the product nature, and those relevant factors should be taken into consideration as well in an e-commerce website evaluation. In addition, the experiment results are dynamic; they are impacted by the network environment. The quantitative mappings might not be available temporarily for those popular websites due to heavy network traffic sometimes, and those popular websites are likely the websites with high scores. *Industrial Equipment* websites are not as popular as book websites or entertainment websites, so they are less impacted by network traffic; on the other hand, book websites or entertainment websites might get lower scores than their real scores due to network traffic, this issue should be addressed in the future experiment.

## 5. SUMMARY AND FUTURE WORK

This paper presented a pilot study on developing an instrument to predict the quality of e-commerce websites. The objective is to provide a meaningful estimation of a given e-commerce website. The 8C model was adopted as the reference model of the heuristic evaluation. Each dimension of the 8C was mapped into quantitative elements by means of web scraping. A software was developed in PHP for both training and testing experiments. 10 experiments were conducted and quantitative analysis was regressively conducted between the experiments. The conversion rate was used to test and verify the heuristic evaluation. It was observed that the trends of the curve for the differences between the expected and actual outcomes was going down along the experiments for both of the testing data and verifying data. This suggested that the mapping revisions between the experiments improved the performance of the evaluation instrument, therefore the experiment process and the revision guideline proposed in Section 2 were on the right track.

However, there are limitations in this study. The experiments only had been done on the home page of each website, although home page is very important for a website and it can provide rich information about the website, it is not sufficient for an e-commerce website, in some cases, the shopping cart or product list are not on the home page. Due to technique incapacity, not all the website features can be mapped into quantitative elements. The experiment results could be impacted by the network environment although that impact is not significant.

The above should be considered in the future work. In addition to that, the mapping revision process could be more robotic by improving the revision guideline (algorithm), for example, the standard deviation for each dimension over all the training websites could be considered in the review process after each experiment in the future work. The evaluation framework should not be limited to the 8C model; it could be extended to include other factors. [16] proposed a number of ways to improve the CR of an e-commerce website, which should be considered in the future study.

Furthermore, for the outcome software instrument, only the scores of each dimension in the 8C model and the total score were output as the website evaluation results. An automatic analysis on the scores could be done and an analysis report could be produced in the future to make the results more comprehensive and helpful.

## REFERENCES

[1]    Li, Fangyu, & Yefei Li, (2011) "Usability evaluation of e-commerce on B2C websites in China.", Procedia Engineering 15 pp5299-5304.

[2]    Bezes, Christophe (2009) "E-commerce website evaluation: a critical review."

[3]    Rayport, Jeffrey F., & Bernard J. Jaworski (2002) Introduction to e-commerce. McGraw-Hill/Irwin marketspaceU.

[4]    Yang, T. Andrew, Dan J. Kim, Vishal Dhalwani, & Tri K. Vu, (2008) "The 8C framework as a reference model for collaborative value Webs in the context of Web 2.0." In Hawaii International Conference on System Sciences, Proceedings of the 41st Annual, pp319-319. IEEE.

[5]    González, Marta, Llúcia Masip, Antoni Granollers, & Marta Oliva, (2009) "Quantitative analysis in a heuristic evaluation experiment." Advances in Engineering Software 40, no. 12. pp1271-1278.

[6]    Hamidizadeh, Mohammad R., Mohammad E. Fadaeinejad, & Fayegh Mojarrad, (2011) "Design of internet marketing based on 7Cs model." In 2011 International Conference on Social Science and Humanity.

[7]    Janssen, Frederik & Johannes Fürnkranz, (2011) "Heuristic rule-based regression via dynamic reduction to classification." In IJCAI Proceedings-International Joint Conference on Artificial Intelligence, vol. 22, no. 1, pp1330.

[8]    Acharya, Ram N., Albert Kagan, Srinivasa Rao Lingam & Kevin Gray, (2011) "Impact Of Website Usability On Performance: A Heuristic Evaluation Of Community Bank Homepage Implementation." Journal of Business & Economics Research (JBER) 6, no. 6.

[9]    Mack, Robert L., and Jakob Nielsen (Eds), (1994) "Usability inspection methods." New York: Wiley & Sons.

[10]   Nielsen, Jakob, and Rolf Molich, (1990) "Heuristic evaluation of user interfaces." In Proceedings of the SIGCHI conference on Human factors in computing systems, pp. 249-256. ACM.

[11]   Nielsen, Jakob, (1995) "Severity ratings for usability problems." http://www.useit.com/papers/heuristic/severityrating.html, last accessed 2018/01/12.

[12]   Nielsen, Jakob, (2013) Conversion Rates. https://www.nngroup.com/articles/conversion-rates/, last accessed 2017/06/28.

[13]   Tullis, Thomas & Albert, William. Measuring the user experience. Morgan Kaufmann; 2008.

[14]   Burstein, Daniel. (2015) Ecommerce Research Chart: Industry benchmark conversion rates for 25 retail categories. http://www.marketingsherpa.com/article/chart/conversion-rates-retail-categories/, last accessed 2016/07/29.

[15]   Chopra, Paras. (2010) Top 10 eCommerce Websites (by Conversion Rate). https://vwo.com/blog/top-ecommerce-websites-conversion-rate/, last accessed 2016/07/29.

[16]   Saleh, Khalid. (2017) The Average Website Conversion Rate by Industry. https://www.invespcro.com/blog/the-average-website-conversion-rate-by-industry/, last accessed 2017 /05/29.

**AUTHORS**

**Dr. Xiaosong Li** obtained her PhD (1999) in Computer Science from University of Auckland in New Zealand. Her research interests include Graphical User Interface, E-Commerce Websites, Machine Learning and etc. She joined Unitec in 2002 where she is a Senior Academic Staff Member.

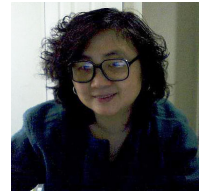